\documentclass[a4paper,11pt]{article}
\pdfoutput=1
\usepackage{appendix}
\usepackage{jcappub}
\usepackage[T1]{fontenc}
\usepackage[style=numeric,backend=biber, sorting = none]{biblatex}

\newcommand{\fnlloc}{f_{\rm NL}^{\rm loc}}
\newcommand{\fnlequil}{f_{\rm NL}^{\rm equil}}
\newcommand{\fnlortho}{f_{\rm NL}^{\rm ortho}}
\newcommand{\omegam}{\Omega_{\rm m}}
\newcommand{\omegab}{\Omega_{\rm b}}
\newcommand{\ns}{n_{\rm s}}
\newcommand{\mmin}{M_{\rm min}}
\newcommand{\invhMpc}{h^{-1}\rm{Mpc}}
\newcommand{\invhGpc}{h^{-1}\rm{Gpc}}
\newcommand{\hinvMpc}{h\rm{Mpc}^{-1}}
\newcommand{\solmass}{h^{-1}M_{\odot}}
\newcommand{\Rs}{R_{\rm s}}

\title{\boldmath Constraining Primordial Non-Gaussianity with Density-Split Clustering}

\author[a,b]{James Morawetz}
\author[a,b,c]{Enrique Paillas}
\author[a,b,d]{Will J. Percival}

\affiliation[a]{Waterloo Centre for Astrophysics, University of Waterloo, 200 University Ave W, Waterloo, ON N2L 3G1, Canada}
\affiliation[b]{Department of Physics and Astronomy, University of Waterloo, 200 University Ave W, Waterloo, ON N2L 3G1, Canada}
\affiliation[c]{Department of Astronomy/Steward Observatory, University of Arizona, 933 North Cherry Avenue, Tucson, AZ 85721, USA}
\affiliation[d]{Perimeter Institute for Theoretical Physics, 31 Caroline St. North, Waterloo, ON N2L 2Y5, Canada}

\emailAdd{jgmorawe@uwaterloo.ca}

\abstract{Obtaining tight constraints on primordial non-Gaussianity (PNG) is a key step in discriminating between different models for cosmic inflation. The constraining power from large-scale structure (LSS) measurements is expected to overtake that from cosmic microwave background (CMB) anisotropies with the next generation of galaxy surveys including the Dark Energy Spectroscopic Instrument (DESI) and Euclid. We consider whether Density-Split Clustering (DSC) can help improve PNG constraints from these surveys for local, equilateral and orthogonal types. DSC separates a surveyed volume into regions based on local density and measures the clustering statistics within each environment. Using the Quijote simulations and the Fisher information formalism, we compare PNG constraints from the standard halo power spectrum, DSC power spectra and joint halo/DSC power spectra. We find that the joint halo/DSC power spectra outperform the halo power spectrum by factors of $\sim$ 1.4, 8.8, and 3.6 for local, equilateral and orthogonal PNG, respectively. This is driven by the higher-order information that DSC captures on small scales. We find that applying DSC to a halo field does not allow sample variance cancellation on large scales by providing multiple tracers of the same volume with different local PNG responses. Additionally, we introduce a Fourier space analysis for DSC and study the impact of several modifications to the pipeline, such as varying the smoothing radius and the number of density environments and replacing random query positions with lattice points.}

\begin{document}
\maketitle
\flushbottom

\section{Introduction}
\label{sec:intro}

The theory of \textit{cosmic inflation} \cite{Guth_1982, Hawking_1982} postulates a brief period of accelerated expansion in the very early universe. This would have driven the universe toward the finely tuned initial conditions necessary to resolve the flatness and horizon problems. It also conveniently provides a mechanism for quantum fluctuations to stretch to macroscopic scales, seeding the initial density perturbations responsible for the large-scale structures we observe today. For these reasons, inflation is now a widely accepted framework among cosmologists. However, we cannot directly observe inflation and the exact mechanism driving it remains poorly understood. Accordingly, cosmologists must search for observational signatures to discriminate among competing theoretical models. See \cite{Achúcarro_2022} for a review of the current state of research on inflation.

Among such observational signatures is the existence, or lack of existence, of \textit{primordial non-Gaussianity} (PNG), typically quantified by the dimensionless parameter $f_{\rm NL}^{\rm X}$. PNG is commonly classified by various shapes (denoted by $\rm X$), each corresponding to a different class of inflation models \cite{Planck_2020} which leave distinct signatures in the primordial bispectrum. While the simplest models predict only small levels of PNG, models that depart from standard slow-roll, single-field assumptions -- e.g., multiple fields, higher-derivative interactions, deviations from the initial Bunch-Davies vacuum, etc. -- can induce large levels of PNG \cite{Desjacques_2010}.

The measurements of cosmic microwave background (CMB) anisotropies from the Planck collaboration \cite{Planck_2020} provide the tightest constraints to date for PNG of local, equilateral and orthogonal types: $ \fnlloc = -0.9 \pm 5.1, \ \fnlequil = -26 \pm 47, \ \fnlortho = -38 \pm 24$. However, the information content from CMB anisotropies is nearly saturated, and large-scale structure (LSS) measurements are soon expected to overtake the CMB in constraining power; next-generation galaxy surveys like the Dark Energy Spectroscopic Instrument (DESI) and Euclid \cite{DESI_2016, Euclid_2011} will probe much larger volumes at higher number densities than previously achieved. Constraining PNG using LSS measurements is challenging, given that non-linear structure growth in the late-time universe inherently leads to non-Gaussianity. This process manifests as a signal in the bispectrum that is degenerate with PNG, making it necessary to identify unique observational signatures to distinguish the two. Local PNG also induces a characteristic scale-dependent bias  (proportional to $1/k^2$ on large scales) for biased tracers of the matter field \cite{Dalal_2008}. Having multiple tracers with different responses to $\fnlloc$, particularly including a single tracer with zero bias \cite{Castorina_2018}, probing the same underlying volume can overcome sample variance and obtain constraints only limited by shot noise \cite{Seljak_2009}. On the other hand, equilateral and orthogonal PNG lack this strong scale-dependent bias contribution, making the power spectrum less informative compared to the bispectrum \cite{Coulton_2023b}.

Many statistics have been proposed to measure higher-order galaxy clustering information. The most straightforward examples include directly measuring the three-point correlation function/bispectrum and four-point correlation function/trispectrum \cite{Sefusatti_2006, Gil-Marin_2017, Slepian_2017, Philcox_2021}. Indirect methods for capturing this higher-order information include counts-in-cells statistics \cite{Szapudi_2004, Uhlemann_2018, Friedrich_2020}, non-linear transformations of the density field \cite{Neyrinck_2009, Wang_2024a, Wang_2024b}, void statistics \cite{Lavaux_2012, Nadathur_2020, Contarini_2023, Fraser_2024}, separate universes \cite{Wagner_2015, Chiang_2015}, marked correlations \cite{White_2016, Massara_2018}, k-th nearest neighbours statistics \cite{Banerjee_2021, Yuan_2024, Coulton_2024}, Minkowski functionals \cite{Schmalzing_1996, Lippich_2021, Appleby_2022, Liu_2023}, the wavelet scattering transform \cite{Valogiannis_2022a, Peron_2024}, skew spectra \cite{Hou_2023}, the minimum spanning tree \cite{Naidoo_2022}, and critical points \cite{Shim_2024}. Several of these clustering methods were compared in a previous mock challenge \cite{Beyond2pt_2024}, and research is ongoing to further understand the complementarity of these different summary statistics.

Among these novel techniques is \textit{Density-Split Clustering} (DSC) \cite{Paillas_2021, Paillas_2023, Paillas_2024, CuestaLazaro_2024}, which measures three-dimensional clustering statistics in separate environments split according to local galaxy density. A similar approach was first applied to cosmic shear analyses from weak gravitational lensing \cite{Gruen_2018, Friedrich_2018, Burger_2023}. The notion of studying galaxy clustering as a function of density environment has been explored in previous works \cite{Abbas_2007, Tinker_2007, Bayer_2021, Bonnaire_2022, Bonnaire_2023}. Most recently, \cite{Paillas_2023} used the Fisher information formalism to study the constraining power of DSC compared to the two-point correlation function (2PCF) for the $\nu \Lambda$CDM model. DSC was also applied to observational data from the CMASS BOSS galaxy sample \cite{Paillas_2024} using an emulator \cite{CuestaLazaro_2024}.

DSC has great potential for constraining PNG for several reasons. First, it captures higher-order information, which is particularly relevant for equilateral and orthogonal PNG, where much of the available information is contained in the bispectrum. Second, the DSC quantiles trace out the matter field on large scales with a wide range of linear biases, including the intermediate quantile that has close to zero bias \cite{Castorina_2018}. If these have different responses to $\fnlloc$ this could, in principle, permit sample variance cancellation to improve local PNG measurements \cite{Dalal_2008, Seljak_2009}. To test both of these hypotheses, we follow the approach of \cite{Paillas_2023} and apply the Fisher information formalism to quantify the information content of DSC statistics when applied to various forms of PNG, marginalizing over cosmological parameters. In doing so, we also introduce a DSC analysis in Fourier space, where power spectra are used as summary statistics, in contrast to previous DSC analyses \cite{Paillas_2021, Paillas_2023, Paillas_2024, CuestaLazaro_2024} that have worked exclusively in configuration space using correlation functions. We also study the impact of varying hyperparameters of the DSC pipeline, such as the smoothing radius, the number of density environments and the type of query positions, all of which may affect information content.

This paper is organized as follows. In section~\ref{sec:sims}, we discuss the simulations used in the analysis and their relevant properties. In section~\ref{sec:method}, we introduce the various summary statistics and describe our methodology. In section~\ref{sec:constraints}, we present our Fisher constraints and discuss modifications to the DSC pipeline. In section~\ref{sec:discuss}, we interpret the results and discuss their ramifications. Finally, in section~\ref{sec:summary}, we conclude with a summary of our findings. All results and data throughout this paper can be reproduced using publicly available codes \footnote{\url{https://github.com/jgmorawetz/densitysplit_fisher_fNL}}.

\section{Simulations}
\label{sec:sims}

In this analysis, we use the \texttt{Quijote} simulations, a suite of N-body simulations designed to quantify the information content of cosmological observables and train machine learning algorithms \cite{VillaescusaNavarro_2020}. Spanning the $(\omegam, \omegab, h, \ns, \sigma_8, M_{\nu}, w)$ hyperplane, they contain 15000 realizations of the fiducial cosmology and 500 realizations of surrounding cosmologies where the parameters are individually varied above and below the fiducial values. Each simulation covers a $1 \ (\invhGpc)^3$ volume, and halo catalogues are constructed using a Friends-of-Friends algorithm \cite{Davis_1985}. We restrict our focus to halos at redshift snapshot $z=0$, and for simplicity, use halos in place of galaxies. A halo mass cut of $\mmin=3.2 \times 10^{13} \  \solmass$ is applied, imposing a mean halo number density of $\bar{n} = 1.55 \times 10^{-4} \ (\hinvMpc)^3$. We also apply separate mass cuts of $\mmin = 3.1 \times 10^{13}\ \solmass,\  3.3 \times 10^{13}\ \solmass$ to marginalize over the parameter $\mmin$ in our Fisher analysis, acting as a proxy for bias parameters \cite{Coulton_2023b}. All measurements are performed in redshift space, where the halo positions are perturbed based on their peculiar velocities along the line-of-sight (LOS) direction. For each of the 500 paired simulations, we generate mocks using the $x$, $y$ and $z$ LOS directions; while the resulting catalogues are correlated, averaging over them helps to reduce noise on derivative estimates. We only use the $z$ LOS direction for computing covariance matrices using the 15000 fiducial simulations.

The recently released \texttt{Quijote-PNG} simulations are an extension of the original set, run with the same code and settings, which probe the effects of various types of PNG on cosmological observables \cite{Coulton_2023a}. The 500 paired simulations for local, equilateral and orthogonal shapes have variations $f_{\rm NL}^{\rm X} = \pm 100$, where $\rm{X}$ is the PNG shape being considered, with the other PNG amplitudes set to 0 and $\Lambda$CDM parameters held at their fiducial values. We restrict our focus to the following parameter space $(\fnlloc, \fnlequil, \fnlortho, \mmin, h, \ns, \omegam, \sigma_8)$ to maintain consistency with other literature \cite{Coulton_2023b, Peron_2024}. Table~\ref{tab:sim_params} lists the parameters associated with each cosmology.

\begin{table}
    \centering
    \begin{tabular}{c | c | c | c | c | c | c | c | c}
     Variation & $\fnlloc$ & $\fnlequil$ & $\fnlortho$ & $\mmin \  (\solmass)$ & $h$ & $\ns$ & $\omegam$ & $\sigma_8$ \\ \hline\hline
     Fiducial & 0  & 0 & 0 & $3.2 \times 10^{13}$ & 0.6711 & 0.9624 & 0.3175 & 0.834 \\
     $f_{\rm NL}^{\rm{loc}, +}$ & 100  &  &  &  &  &  &  & \\
     $f_{\rm NL}^{\rm{loc}, -}$ & -100  &  &  &  &  &  &  &  \\
    $f_{\rm NL}^{\rm{equil}, +}$ &  & 100 &  &  &  &  &  & \\
    $f_{\rm NL}^{\rm{equil}, -}$ &  & -100 &  &  &  &  &  & \\
    $f_{\rm NL}^{\rm{ortho}, +}$ &   &  & 100 &  &  &  &  & \\
    $f_{\rm NL}^{\rm{ortho}, -}$ &   &  & -100 &  &  &  &  & \\
    $M_{\rm min}^+$ &  &  &  & $3.3 \times 10^{13}$ &  &  &  &  \\
    $M_{\rm min}^-$ &  &  &  & $3.1 \times 10^{13}$ &  &  &  &  \\
    $h^+$ &  &  &  &  & 0.6911 &  &  &  \\
    $h^-$ &  &  &  &  & 0.6511 &  &  &  \\
    $n_{\rm s}^+$ &  &  &  &  &  & 0.9824 &  &  \\
    $n_{\rm s}^-$ &  &  &  &  &  & 0.9424 &  &  \\
    $\Omega_{\rm m}^+$ &  &  &  &  &  &  & 0.3275 &  \\
    $\Omega_{\rm m}^-$ &  &  &  &  &  &  & 0.3075 &  \\
    $\sigma_8^+$ &  &  &  &  &  &  &  & 0.849 \\
    $\sigma_8^-$ &  &  &  &  &  &  &  & 0.819 \\
    \end{tabular}
\caption{The parameter values for each of the cosmologies being studied. Apart from the top row, only values that differ from the fiducial choices are shown. \label{tab:sim_params}}
\end{table}

\section{Methodology}
\label{sec:method}

\subsection{Density-Split Clustering (DSC)}

The DSC procedure divides a sample volume into separate regions based on local galaxy density and measures the clustering statistics of each environment. We use the publicly available pipeline\footnote{\url{https://github.com/epaillas/densitysplit}} and briefly summarize the steps and modifications needed for our analysis:

\begin{enumerate}

\item{Paint the redshift-space halo positions onto a mesh grid of dimension $N_{\rm mesh}$ covering the sample volume using the Triangular Shaped Cloud (TSC) window scheme, and smooth the resulting field (in Fourier space) with a Gaussian filter of radius $\Rs$.}

\item{Fill the sample volume with a large number of query points $N_{\rm query}$, and measure the smoothed density contrast $\Delta(\Rs)$ at each point. Existing DSC analyses use randomly placed query positions (specifically, five times as many randoms as halos $N_{\rm query} = 5N_{\rm halo}$). For reasons discussed later, we instead use equally spaced lattice points such that each voxel in the grid has one query point at its centre, giving $N_{\rm query} = N_{\rm nmesh}^3$.}

\item Split the query points into separate quantiles based on overdensity. The existing convention is to use five bins, or quintiles, in total.

\item Measure clustering statistics of the quantiles individually and concatenate them into a single data vector to constrain cosmological parameters. This includes the cross-correlation of the quantiles with the halos (in redshift space) and the autocorrelation of the quantiles.

\end{enumerate}

\begin{figure}
    \centering
    \includegraphics[width=\textwidth]{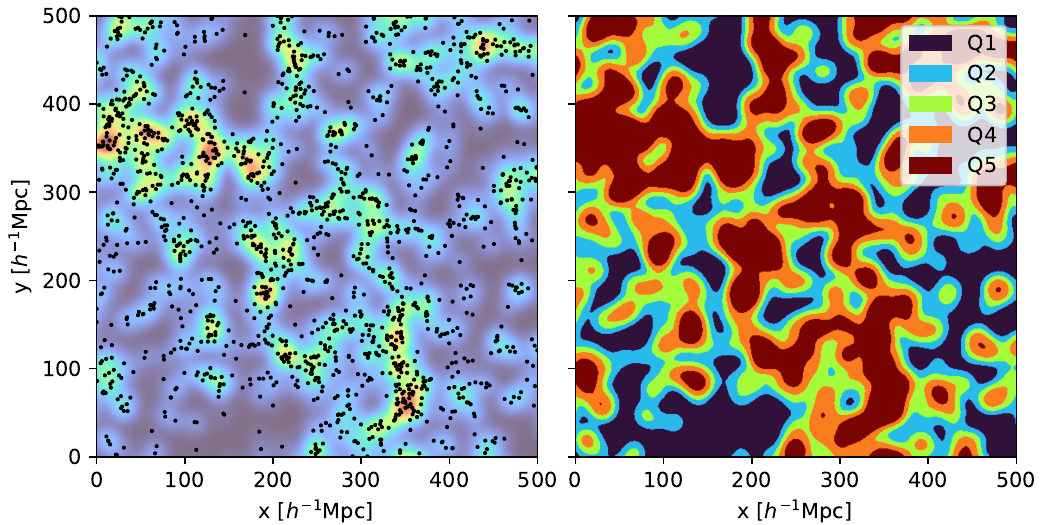}
    \caption{A $500 \times 500 \ \invhMpc$ cross-section for one of the Quijote fiducial simulation volumes. The slice depth is set to twice the smoothing radius. The left panel shows the projected halo positions in redshift space, and the background colour represents the smoothed density contrast. The right panel shows a map of the associated quantiles. Q1 denotes the most underdense quantile, Q5 denotes the most overdense quantile, and Q2-4 are the intermediate density quantiles.}
    \label{fig:ds_cross_section}
\end{figure}

To help visualize, Fig.~\ref{fig:ds_cross_section} depicts an example cross-sectional map demonstrating the DSC methodology applied to one of the Quijote fiducial simulation volumes. The existing DSC Fisher analysis \cite{Paillas_2023} used a TopHat filter of radius $\Rs = 20 \ \invhMpc$ and performed smoothing in configuration space. We instead follow the convention of more recent DSC analyses \cite{Paillas_2024, CuestaLazaro_2024} and use a Gaussian filter of radius $\Rs = 10 \ \invhMpc$, corresponding to an effective TopHat radius (with asymptotically converging window functions in Fourier space as $k \rightarrow 0$) of $\Rs = 10 \sqrt{5} \ \invhMpc \sim 22.4 \ \invhMpc$, and perform the smoothing in Fourier space to avoid the computationally expensive pair counting procedure. However, instead of fixing the cell size to $5 \ \invhMpc$ for the mesh grid, we use a resolution of $N_{\rm mesh} = 512$, which, for the given box size, corresponds to a cell size of $\sim 2 \ \invhMpc$, helping to reduce additional smoothing effects beyond the Gaussian filter. Existing DSC analyses use the Cloud-In-Cell (CIC) interpolation scheme to paint halos to the mesh, but we instead employ the TSC scheme to further reduce aliasing effects. We do not apply the interlacing technique \cite{Sefusatti_2016} since it significantly increases computational resources. Lastly, the halo positions are considered in redshift space, which can effectively blur out some of the local density information. The first DSC Fisher analysis \cite{Paillas_2023} showed that reconstruction algorithms can generate a pseudo real-space catalogue to recapture some of the lost information, but also showed that it can introduce additional biases. We choose not to apply reconstruction here and accept the small information loss to avoid unwanted effects and more accurately mimic the conditions of a true galaxy survey.

\subsection{Power Spectrum Estimation}
\label{sec:power}

The existing DSC analyses \cite{Paillas_2021, Paillas_2023, Paillas_2024, CuestaLazaro_2024} work exclusively in configuration space, i.e., using correlation functions as summary statistics. While the power spectrum and correlation function are a Fourier transform pair and encode the same information, there can be significant differences in computational resources and constraining power depending on the scales being probed and the number densities involved. For example, the dominant contribution to computing time for Fourier space is the Fast Fourier Transform (FFT) procedure, which depends only on the resolution of the grid; painting the particles to the mesh is typically a subdominant contribution. On the other hand, configuration space involves a pair counting procedure that is highly sensitive to the number density of the sample. We introduce a DSC Fourier space analysis, which could prove useful for DSC analyses on high-density targets, such as the DESI Bright Galaxy Sample \cite{Hahn_2023}. We leave a more direct comparison between the constraining power and computational resources of both approaches to future work.

To maintain consistency, we use the same grid resolution and resampling and interlacing schemes for the computation of the power spectra as when applying the density-split step to the halo field, except the halo power spectrum where we use CIC interpolation since it is used in other literature \cite{Coulton_2023b}. The power spectrum computations were implemented using the publicly available packages \texttt{pypower}\footnote{\url{https://github.com/cosmodesi/pypower}} and \texttt{nbodykit}\footnote{\url{https://github.com/bccp/nbodykit}}. The minimum wavenumber is set to the fundamental mode $k_{\rm min} = 2 \pi/L$ where $L = 1 \ \invhGpc$ is the box size, and the maximum wavenumber is set to the Nyquist mode $k_{\rm max} = \pi N_{\rm mesh}/L$, although we limit our analysis to scales $k \leq 0.5 \ \hinvMpc$ again for consistency with other literature \cite{Coulton_2023b, Peron_2024}. The bin width is set equal to the minimum wavenumber. The multipole moments of the cross power spectrum between fields $\mathrm{a}$ and $\mathrm{b}$ are calculated by taking the real component of: 
\begin{equation}\label{eq:2} 
    \hat{P}_l^{\mathrm{ab}}(k_{\rm i}) = \frac{2l+1}{2N_{\rm i}} \sum_{k_{\rm i, min}  \ \leq  \ |\vec{k}| \ \leq \ k_{\rm i, max}} L_l(\mu)  \delta^{\mathrm{a}}(\vec{k})\delta^{\mathrm{b}{\star}}(\vec{k}) \,,
\end{equation} 
where $\rm{i}$ denotes the $k$ bin index with $k_{\rm i, max} - k_{\rm i, min} = \Delta k$ being the bin width. $N_{\rm i}$ is the normalization, $\delta(\vec{k})$ are the Fourier modes of the overdensity field and $L_l(\mu)$ is the Legendre polynomial of order $l$, where $l=0, 2$ correspond to the monopole and quadrupole moments, respectively\footnote{We only compute the monopole and quadrupole moments of the power spectra; the hexadecapole also contains non-trivial information in redshift space, but it is too noisy to accurately compute derivatives and contains comparatively little information.}. We also subtract the $1/\bar{n}$ shot noise contribution for the halo auto power spectra, but not for the quantile auto power spectra for reasons discussed later.

The relevant summary statistics include the quantile auto power spectra, denoted $P^{\rm qq}_{(0, 2)}$, the quantile-halo cross power spectra, denoted $P^{\rm qh}_{(0, 2)}$, and the halo auto power spectra, denoted $P^{\rm hh}_{(0, 2)}$, where $\rm{q}$ denotes the quantile index, $\rm{h}$ denotes the halos and the subscripts $(0, 2)$ denote the monopole and quadrupole moments. The previous DSC Fisher analysis \cite{Paillas_2023} focused exclusively on the information content of DSC compared to the standard 2PCF. In our analysis, we also study the joint constraining ability of the DSC power spectra and the halo power spectrum. 

\begin{figure}
    \centering
    \includegraphics[width=\textwidth]{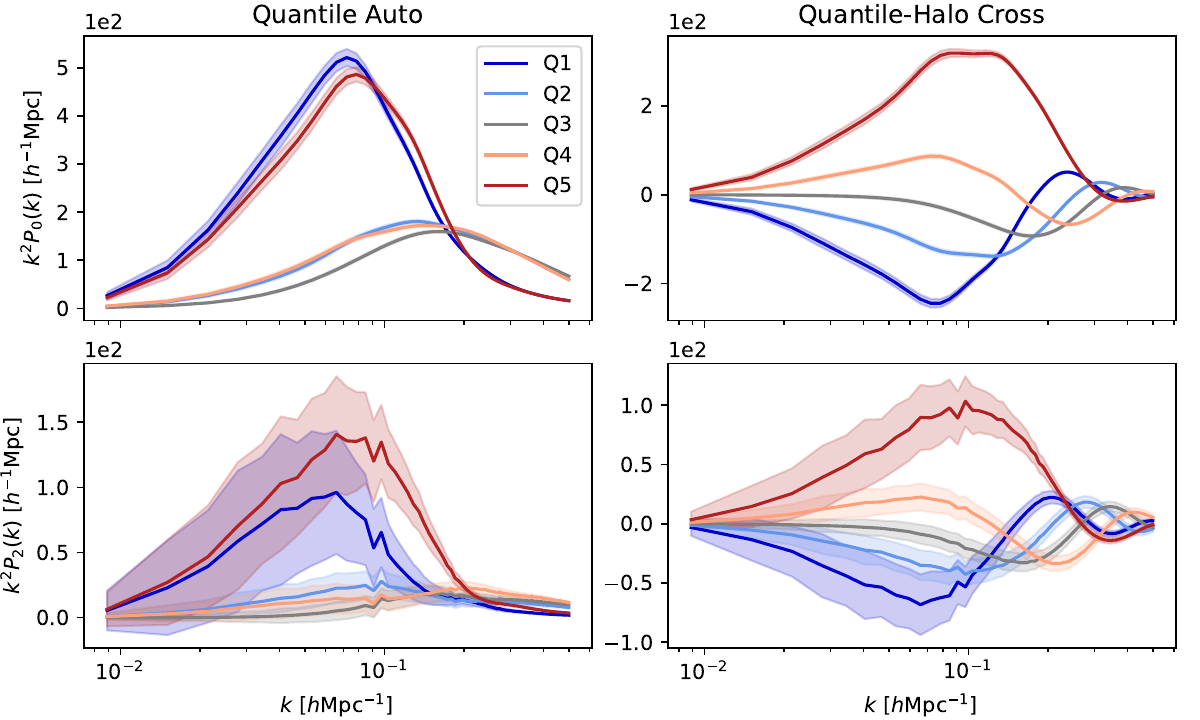}
    \caption{The DSC power spectra for the fiducial cosmology. The solid line is the mean across all realizations, while the shaded region denotes the standard noise for a single realization. The top (bottom) panel denotes the monopole (quadrupole) moment, and the left (right) panel denotes the quantile auto (quantile-halo cross) power spectra.}
    \label{fig:fiducial_power}
\end{figure}

Fig.~\ref{fig:fiducial_power} shows the monopole and quadrupole of the DSC auto and cross power spectra for the fiducial cosmology. For the auto power spectra on the left, we see the expected behaviour caused by scale-independent bias on large scales and a rapid decrease in power on small scales caused by the Gaussian filter (this suppression of power on small scales is more pronounced compared to the cross power spectra where the halo field is not smoothed). For the cross power spectra on the right, the large-scale behaviour is also driven by the different linear biases of the quantiles and halos. But unlike the auto power spectra which are strictly positive, the negatively- and positively-biased quantile responses have opposite signs on large scales and exhibit visible symmetry around the horizontal axis. On small scales, the functions have a zero-crossing and a series of peaks and troughs representing characteristic scales at which the density modes of the quantile and halo fields are correlated (peaks) or anti-correlated (troughs); these scales are directly controlled by the choice of smoothing radius. Moreover, the density distribution is highly skewed on small scales (i.e., the quantile overdensity boundaries are not symmetric around $\delta=0$), meaning these small-scale features occur at different scales for each quantile.

Given that the query positions – lattice points or randoms – are distributed uniformly before splitting into quantiles, the overdensity fields by definition sum to zero. In other words: 
\begin{equation}\label{eq:3} 
    \delta_{\rm i}(\vec{x}) = -\sum_{\rm j \neq i} \delta_{\rm j}(\vec{x}) \,,
\end{equation} 
where $\rm{i}$ and $\rm{j}$ are indices denoting the quantiles. By linearity of expectations, substituting Eq.~\ref{eq:3} gives the following expression for the quantile-halo cross-correlation: 
\begin{equation}\label{eq:4} 
    \epsilon_{\rm i,h}(\vec{r}) = \left < \delta_{\rm i}(\vec{x})\delta_{\rm h}(\vec{x}-\vec{r}) \right > = -\sum_{\rm j\neq i} \left < \delta_{\rm j}(\vec{x})\delta_{\rm h}(\vec{x}-\vec{r})\right > = -\sum_{\rm j \neq i} \epsilon_{\rm j,h}(\vec{r}) \,,
\end{equation} 
where $\rm{h}$ denotes the halos. Because any quantile-halo cross-correlation can be expressed as a sum of the remaining quantile-halo cross-correlations, all information from the quantile-halo cross power spectra is encoded in all but one of the available quantiles. We note however that the equivalent argument does not apply for the quantile autocorrelations. To see why, we compute the quantile autocorrelation: 
\begin{equation}\label{eq:5} 
\epsilon_{\rm i,i}(\vec{r}) = \left < \delta_{\rm i}(\vec{x})\delta_{\rm i}(\vec{x}-\vec{r}) \right > = \sum_{\rm j \neq i} \sum_{\rm k \neq i} \left < \delta_{\rm j}(\vec{x})\delta_{\rm k}(\vec{x}-\vec{r})\right > = \sum_{\rm j \neq i} \epsilon_{\rm j,j}(\vec{r}) + \sum_{\rm j, k \neq i, j \neq k} \epsilon_{\rm j, k}(\vec{r}) \,.
\end{equation} 
Eq.~\ref{eq:5} demonstrates that any quantile autocorrelation is not merely a sum over the other quantile autocorrelations, but also a sum over the quantile-quantile cross-correlations. The latter are not included under the current implementation of DSC, and thus all quantiles must be included to capture all available information from the quantile auto power spectra. We follow the convention of previous DSC studies \cite{Paillas_2023, Paillas_2024, CuestaLazaro_2024} and omit the intermediate quantile Q3 while keeping Q1, Q2, Q4, Q5. But instead of omitting the quantile-halo cross and quantile auto power spectra, we merely omit the quantile-halo cross power spectra. 

As introduced earlier, previous applications of DSC \cite{Paillas_2021, Paillas_2023, Paillas_2024, CuestaLazaro_2024} have used randomly placed query positions for the quantiles. This approach introduces unnecessary shot noise on small scales. One solution is to populate the volume with a sufficiently large number of randoms such that the additional shot noise is negligible – this is computationally demanding since the average number of randoms per pixel in the grid must be significantly larger than one. Instead, assigning a single query point at the centre of each pixel removes the shot noise introduced when using random query positions, leaving only the shot noise due to the discreteness of the underlying halo field.
To test the validity of this alternate approach, appendix~\ref{ap:query_position_types} presents a plot comparing the observed power spectra for both approaches – we find close agreement over a broad range of scales. Nevertheless, small differences can be neglected given that an application to data would likely involve an emulator calibrated directly on simulations that will match either convention. In section~\ref{sec:constraints}, we discuss the differences in constraining power between these two methods.

\subsection{Fisher Information}
\label{sec:fisherinfo}

We employ the Fisher information formalism \cite{Fisher_1935} to quantify the information content of our statistics. Under the simplifying assumption that our data vectors follow multivariate Gaussian distributions and that their covariance matrices do not change significantly with cosmological parameters, the Fisher information matrix simplifies greatly: 
\begin{equation} \label{eq:6} 
    \boldsymbol{F}_{\rm ij}(\boldsymbol{\theta^{\star}}) = \frac{\partial \boldsymbol{\mu}^T}{\partial \theta_{\rm i}}\boldsymbol{C}^{-1}\frac{\partial \boldsymbol{\mu}}{\partial \theta_{\rm j}} \bigg\rvert_{\boldsymbol{\theta^{\star}}}\,,
\end{equation} 
where $\boldsymbol{\mu}$ is the noise-free data vector as a function of parameter vector $\boldsymbol{\theta}$, $\boldsymbol{C}$ is the covariance matrix of the data vector, and $\rm i,j$ represent the ith and jth indices of the Fisher matrix. These quantities are evaluated around the fiducial model vector $\boldsymbol{\theta}^{\star}$. The Cram\'{e}r-Rao bound states that the inverse Fisher matrix provides a lower bound on the parameter variances – equality holds if the optimal estimator is utilized: 
\begin{equation} \label{eq:7} 
\sigma(\theta_{\rm i}) \geq \sqrt{F^{-1}_{\rm ii}} \,.
\end{equation} 
We estimate the derivatives and inverse covariance matrices numerically: 
\begin{equation} \label{eq:8} 
    \frac{\partial \boldsymbol{\mu}}{\partial \theta_{\rm i}}\bigg\rvert_{\boldsymbol{\theta^{\star}}} \approx \frac{\boldsymbol{\bar{\mu}}(\theta^{\star}_{\rm i} + d\theta_{\rm i}) - \boldsymbol{\bar{\mu}}(\theta^{\star}_{\rm i} - d\theta_{\rm i})}{2d\theta_{\rm i}}\,,
\end{equation} 

\begin{equation} \label{eq:9} 
    \boldsymbol{C}^{-1} \equiv \frac{N_{\text{cov}} - N_{\text{bin}} + N_{\theta}-1}{N_{\text{cov}} - 1} \boldsymbol{C'}^{-1}, \ \ \boldsymbol{C'} \equiv \frac{1}{N_{\text{cov}} - 1}\sum_{\rm i}^{N_{\text{cov}}} (\boldsymbol{\mu_{\rm i}}^{\star} - \boldsymbol{\bar{\mu}}^{\star})(\boldsymbol{\mu_{\rm i}}^{\star} - \boldsymbol{\bar{\mu}}^{\star})^T\,,
\end{equation} 
where $N_{\rm cov}$ is the number of simulations used to compute the covariance matrix, $N_{\rm bin}$ is the number of entries in our data vector, and $N_{\theta}$ is the number of parameters included in the Fisher matrix. $\boldsymbol{\mu_{\rm i}}$ is the ith realization of the data vector and $\boldsymbol{\bar{\mu}}$ is the mean data vector across all realizations, for either the covariance matrix or the numerical derivatives. The $\star$ symbol denotes quantities evaluated at the fiducial cosmology. We apply the correction factor from \cite{Percival_2022} to debias our estimate of the inverse covariance matrix – similar to \cite{Hartlap_2007} but also accounting for the inversion of the Fisher matrix.

\section{PNG Fisher Constraints from DSC}
\label{sec:constraints}

Going forward, we shall refer to the following combinations of summary statistics:

$$\vec{d} \equiv \{P^{\rm hh}_{(0, 2)}\} \ \   \rightarrow \ \  \text{`Halo'}$$ $$\vec{d} \equiv \{P^{\rm ii}_{(0, 2)}, P^{\rm jh}_{(0, 2)}\}, {\rm i} = (1,2,3, 4,5), {\rm j} = (1,2,4,5)  \ \  \rightarrow \ \  \text{`DSC'}$$ $$\vec{d} \equiv \{P^{\rm hh}_{(0, 2)}, P^{\rm ii}_{(0, 2)}, P^{\rm jh}_{(0, 2)}\}, {\rm i} = (1,2,3,4,5), {\rm j} = (1,2,4,5) \ \   \rightarrow \ \  \text{`Joint'}$$

\subsection{Raw Constraints}

\begin{figure}
    \centering
    \includegraphics[width=\textwidth]{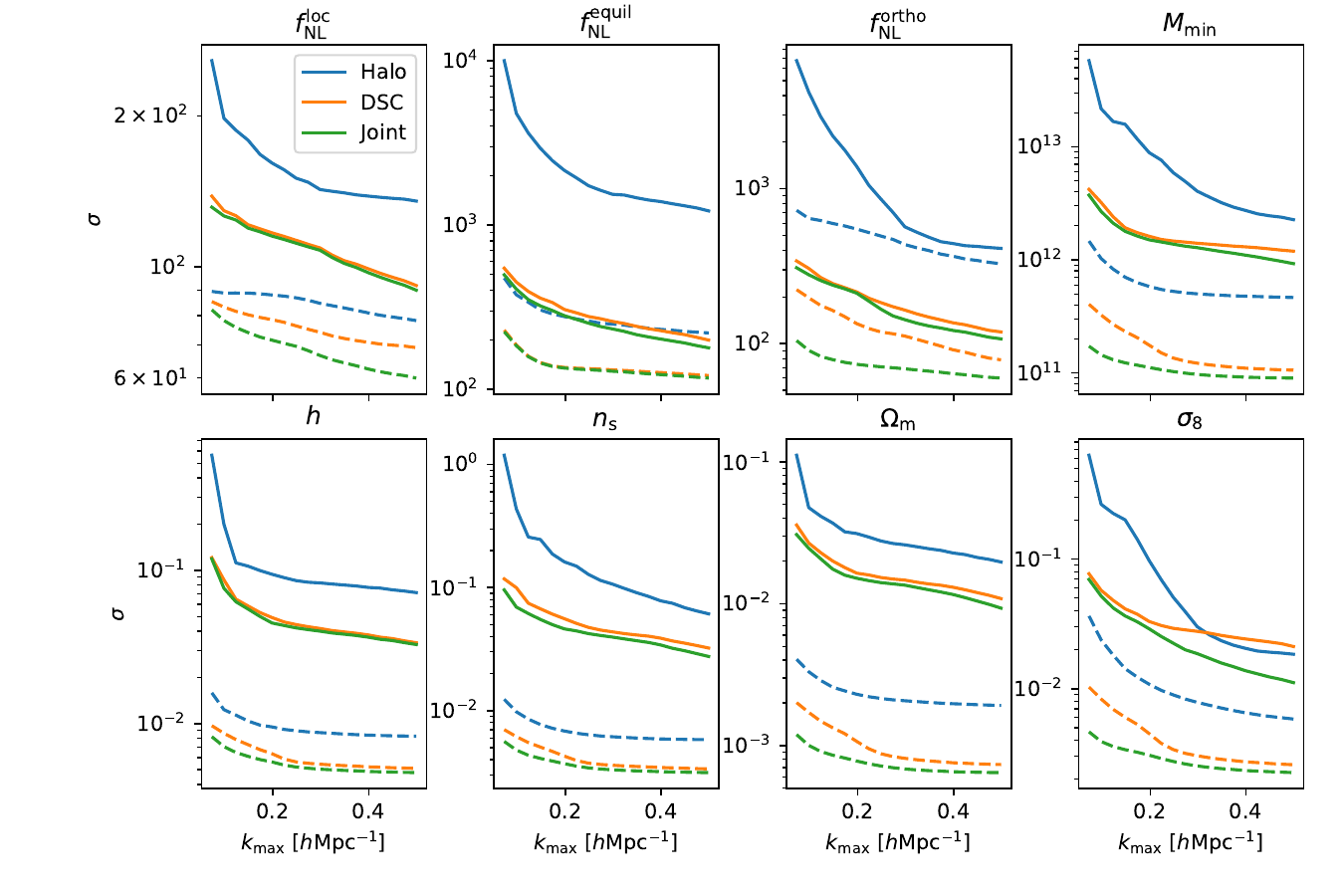}
    \caption{Raw, marginalized (solid line) and unmarginalized (dashed line) constraints, in a $1 \ (\invhGpc)^3$ volume with $\bar{n} = 1.55 \times 10^{-4} \ (\hinvMpc)^3$ at $z=0$, as a function of maximum fitting wavenumber for the halo power spectrum, DSC power spectra or joint power spectra.}
    \label{fig:constraints_max_wavenumber}
\end{figure}

\begin{figure}
    \centering
    \includegraphics[width=\textwidth]{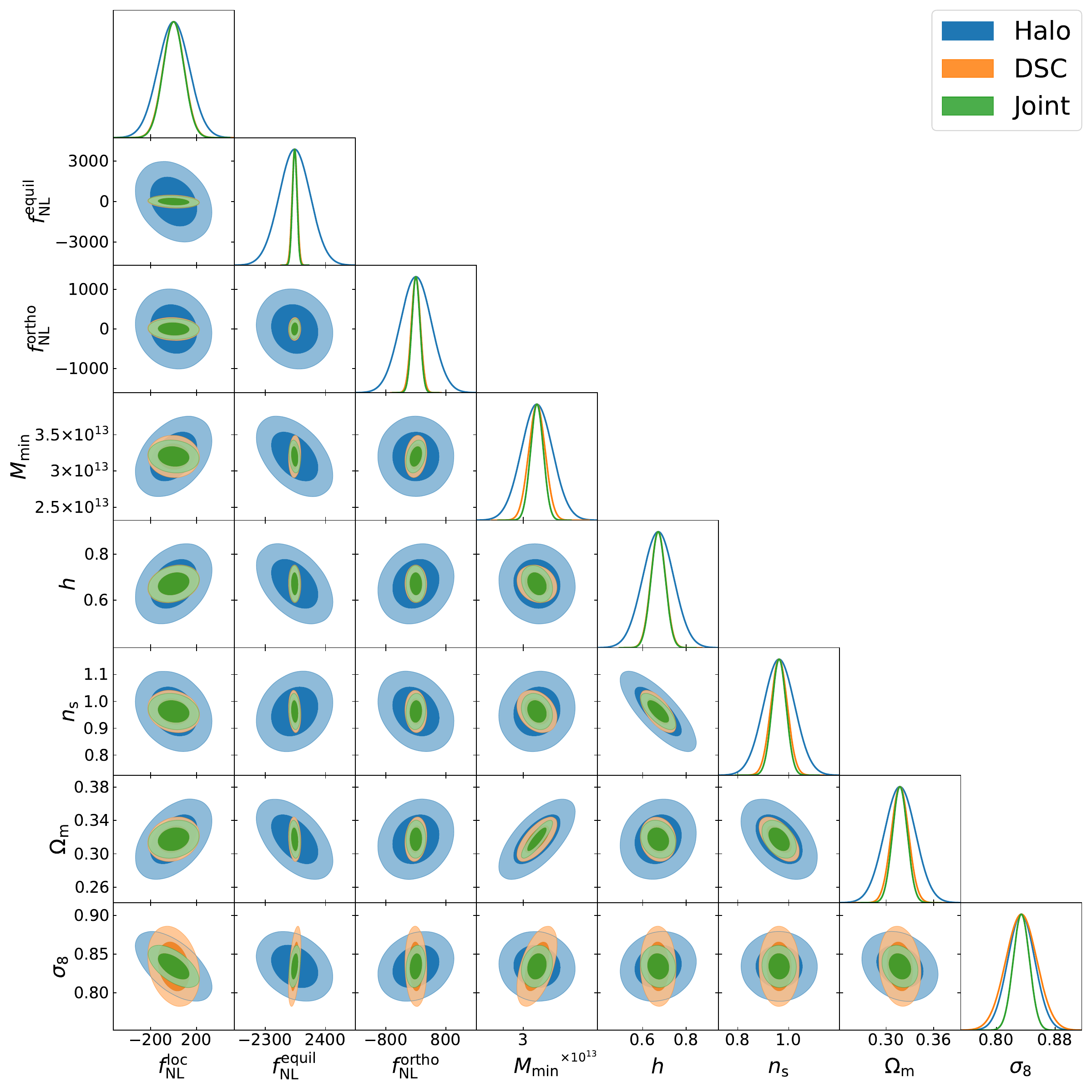}
    \caption{Contours (1$\sigma$ and 2$\sigma$) for the raw, marginalized constraints, in a $1 \ (\invhGpc)^3$ volume with $\bar{n} = 1.55 \times 10^{-4} \ (\hinvMpc)^3$ at $z=0$, with the maximum wavenumber $k_{\rm max}=0.5 \ \hinvMpc$ for the halo power spectrum, DSC power spectra or joint power spectra.}
    \label{fig:constraints_contour}
\end{figure}

\begin{figure}
    \centering
    \includegraphics[width=\textwidth]{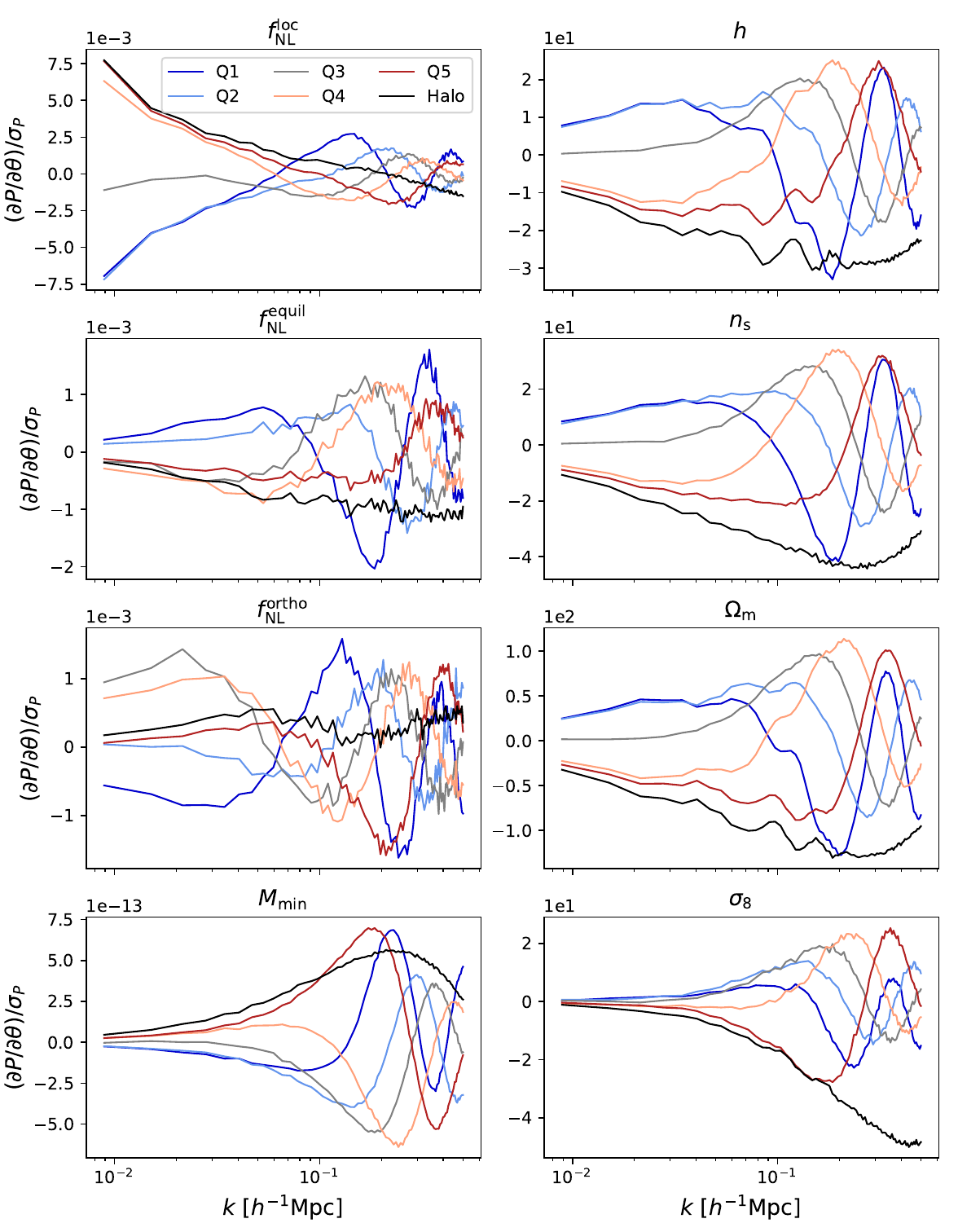}
    \caption{Partial derivatives, in units of the fiducial standard noise, of the DSC quantile-halo cross (and halo auto) power spectrum monopoles with respect to each parameter considered in the analysis. The DSC quantile auto power spectrum monopoles (and the quadrupoles in both cases) are not displayed as they contain more noise and provide less information.}
    \label{fig:monopole_derivatives}
\end{figure}

Fig.~\ref{fig:constraints_max_wavenumber} shows the constraints for each of the three statistics as a function of the maximum wavenumber. This includes the marginalized (over the PNG shapes, $\Lambda$CDM/mass cut parameters) and unmarginalized constraints. As expected, we observe weakened constraints when marginalizing over other parameters. Fig.~\ref{fig:constraints_contour} presents the corresponding constraints at the maximum wavenumber $k_{\rm max}=0.5 \ \hinvMpc$, including parameter covariances. Fig.~\ref{fig:monopole_derivatives} displays the partial derivatives of the DSC quantile-halo cross (and halo auto) power spectrum monopoles with respect to each parameter. The responses are expressed in units of the standard fiducial noise to highlight the relative constraining power provided from different scales. With the exception of $\fnlloc$ (due to scale-dependent bias on large scales), more of the constraining power is found on small scales where non-linearities give rise to higher-order information. Furthermore, each quantile has a distinct sensitivity to the parameters, which is responsible for the degeneracy-breaking capability of DSC.

\subsection{Convergence Corrections}
\begin{figure}
    \centering
    \includegraphics[width=\textwidth]{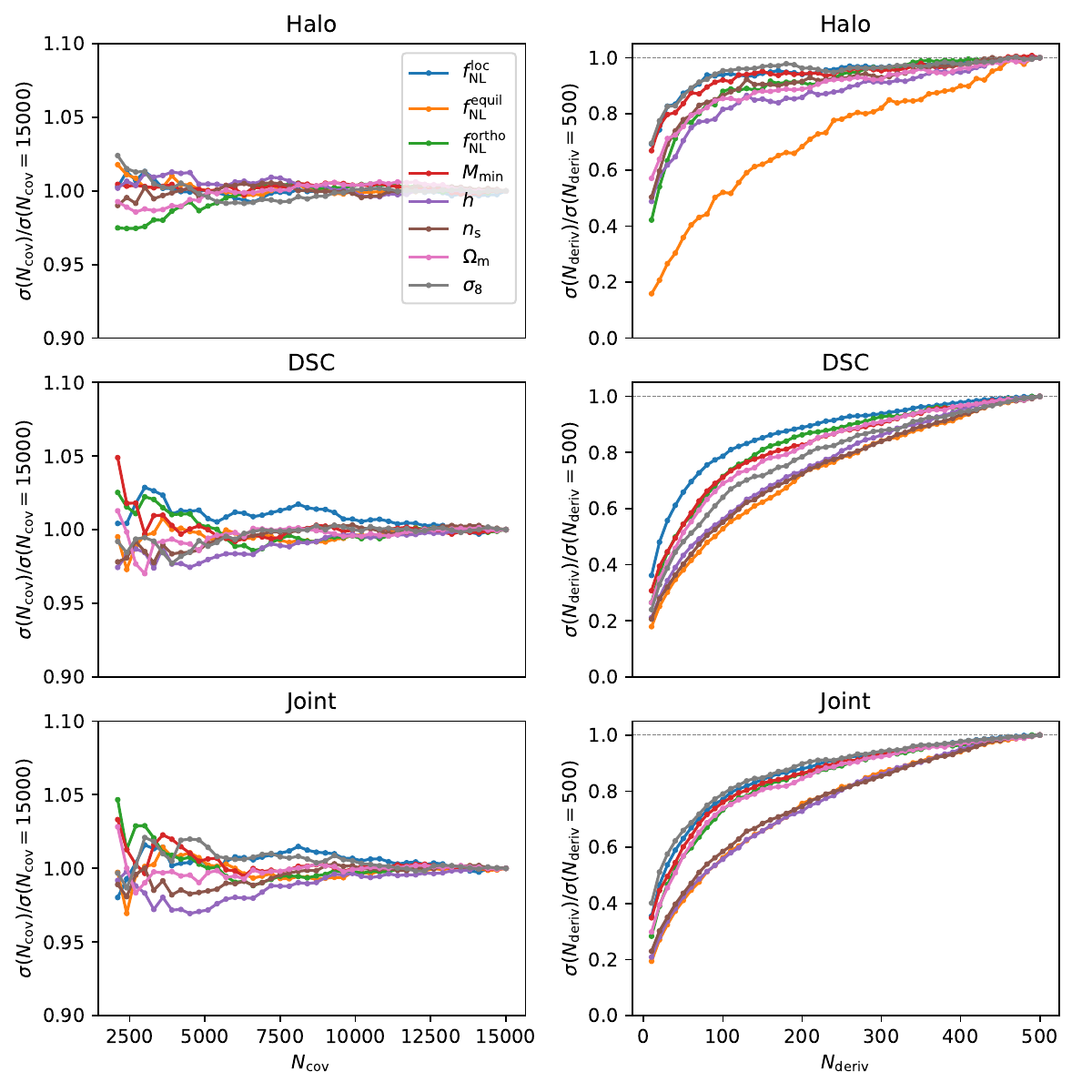}
    \caption{Marginalized constraints normalized to the maximum number of available covariance realizations (left panel) and derivative realizations (right panel).}
    \label{fig:convergence}
\end{figure}

We must test the robustness of these results before drawing conclusions. Because we are averaging over a finite number of realizations to compute covariance matrices and derivatives, the estimated constraints will contain noise. Covariance matrices are estimated using 15000 realizations, significantly larger than the 500 realizations used to estimate derivatives. Fig.~\ref{fig:convergence} shows the marginalized constraints as a function of the number of realizations used to estimate covariance matrices and derivatives, normalized to the maximum number of realizations. Each derivative realization is the average of the three LOS directions, while each covariance realization uses just one LOS direction. We observe only percent-level fluctuation with respect to the number of covariance realizations and do not observe any systematic offsets, given that we have applied the corrections to debias our inverse Fisher matrices \cite{Percival_2022}.

On the other hand, the noisy derivatives lead to systematically over-optimistic constraints. To understand this intuitively, we provide an analytic proof in Appendix~\ref{ap:convergence_issues}; we follow the approach motivated by \cite{Hou_2023} and derive the functional dependence of the constraints on the number of derivative realizations assuming the covariance matrices are known exactly – a reasonable approximation given the large number of realizations compared to the derivatives. The derivation in Appendix~\ref{ap:convergence_issues} leads to the following dependence: 
\begin{equation} \label{eq:10}
    \sigma(\theta_{\rm i})_{\rm corrected} \approx f(\infty) \sigma(\theta_{\rm i})_{\rm raw}, \  f(N_{\rm deriv}) = \left ( \frac{1+C}{1+\frac{500C}{N_{\rm deriv}}}\right)^{1/2} \,,
\end{equation} 
where $C$ is the fitting constant representing the level of convergence and $N_{\rm deriv}$ is the number of derivative realizations. After fitting $C$, we take the limit as $N_{\rm deriv} \rightarrow \infty$ and apply the correction factor $f(\infty) = (1+C)^{1/2}$ to obtain robust estimates. Given that each parameter has a different level of convergence, we fit the expression for each parameter separately. We note that the derived expression only strictly holds in the limit of large $N_{\rm deriv}$, where the central limit theorem applies; we conservatively fit the model only between 100 and 500 derivative realizations to account for this.

\begin{figure}
    \centering
    \includegraphics[width=\textwidth]{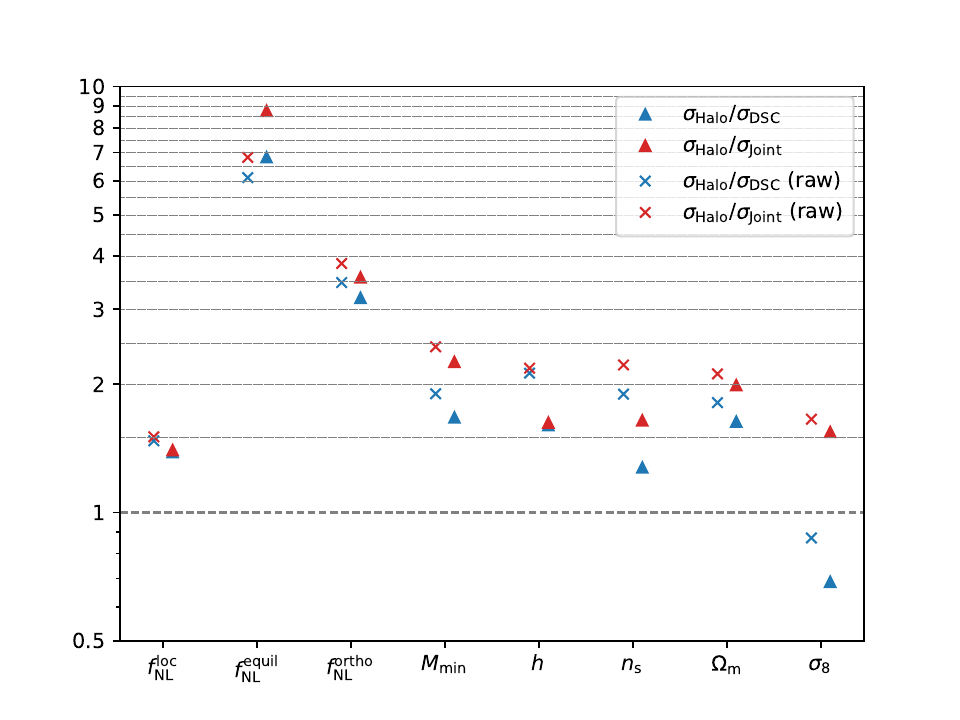}
    \caption{Relative improvement (raw and corrected) for the DSC and joint analyses, compared to the halo-only analysis, across the parameters fitting to the maximum wavenumber of $k_{\rm max} = 0.5 \ \hinvMpc$.}
    \label{fig:raw_corrected_relative}
\end{figure}

The constraints are less strongly converged for the DSC and joint analyses compared to the halo-only analysis. Without applying the proposed corrections above, we would be overestimating the relative improvement from adding DSC functions. Fig.~\ref{fig:raw_corrected_relative} shows the constraints for the DSC and joint fits normalized to the halo-only constraints. The joint analysis provides factors of $\sim$ 1.4, 8.8, 3.6, 2.3, 1.6, 1.6, 2.0, and 1.5 improvement over the halo-only analysis for parameters $\fnlloc, \fnlequil, \fnlortho, \mmin, h, n_s, \omegam, \sigma_8$, after applying convergence corrections. Except $\fnlequil$, the relative improvements for the DSC and joint analyses from the halo-only analysis are weakened upon applying the convergence correction, confirming the importance of applying these corrections to avoid over-optimistic forecasts. The joint analysis provides improvement over the halo-only analysis for all parameters, while the DSC analysis provides improvement for all parameters except $\sigma_8$, highlighting the importance of the joint constraining power of the halo/DSC power spectra.

We emphasize that the constraints presented in this paper should only serve as a comparison of the relative constraining power between DSC and two-point statistics, not as a forecast for future surveys such as DESI. Under more realistic modelling, halos should be populated with galaxies using a model such as the Halo Occupation Distribution (HOD); the constraints would likely weaken upon marginalizing over the model parameters. At the same time, surveys like DESI will probe a significantly larger volume than $1 \ (\invhGpc)^3$ and sample galaxies at higher number density than $1.55 \times 10^{-4} \ (\hinvMpc)^3$, which would improve the constraints. A simulation-based emulator approach (previously applied to CMASS BOSS \cite{Paillas_2024, CuestaLazaro_2024}) would also capture any redshift dependence of our statistics. A rigorous analysis of the constraining power and a comparison to other higher-order alternative clustering methods is left to future work.

\subsection{Modifications to Hyperparameters}

\begin{figure}
    \centering
    \includegraphics[width=\textwidth]{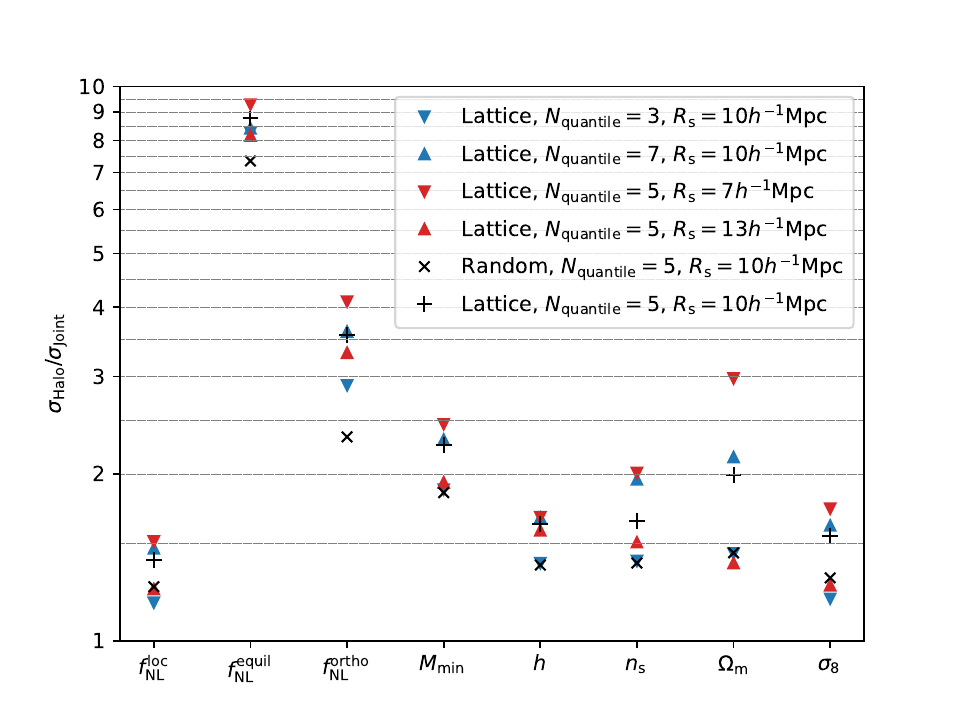}
    \caption{Relative improvement (corrected for derivative convergence) for the joint analysis compared to the halo-only analysis fitting to the maximum wavenumber of $k_{\rm max} = 0.5 \ \hinvMpc$, for variations in the smoothing radius, number of quantiles and query type.}
    \label{fig:hyperparameter_comparisons}
\end{figure}

The main results of this paper, along with existing DSC analysis \cite{Paillas_2024, CuestaLazaro_2024}, use a fixed Gaussian smoothing radius $\Rs = 10 \ \invhMpc$ and number of bins $N_{\rm quantile}=5$. This choice of hyperparameters is not necessarily optimal for extracting maximal information. We additionally test smoothing radii of $\Rs = 7, 13 \ \invhMpc$, number of density environments $N_{\rm quantile} = 3, 7$, and switching from randoms (five times the number of halos) to lattice query points in Fig.~\ref{fig:hyperparameter_comparisons}. 

For reasons discussed in section~\ref{sec:method}, we find that switching from random to lattice query positions boosts constraining power for all parameters, matching the limit of a larger number of random points, and reducing the associated shot noise. Except $\fnlequil$, all constraints improve as the number of quantiles increases. A larger number of density bins more completely characterizes the non-Gaussian density distribution. At the same time, increasing the number of quantiles also increases the computational expense and the size of the data vector – an important trade-off given that a sufficient number of realizations, relative to the size of the data vector, is needed to obtain a reliable estimate of the covariance matrix. We also observe consistent improvement in the constraints when using smaller smoothing radii – excessively large $\Rs$ washes out small-scale information. Since halos are a discrete tracer of the matter field, the smoothing radius must be large enough to ensure a sufficient range of densities and prevent it from being highly skewed. The number densities considered in this paper are low compared to upcoming surveys, which will permit smaller smoothing radii before running into this problem. We are also using halos in place of galaxies, which neglects effects arising from the galaxy-halo connection. 

\section{Interpretation of Results}
\label{sec:discuss}

\subsection{Higher-Order Information}

The most significant improvement between the joint DSC/halo analysis and halo-only analysis occurs for PNG of equilateral and orthogonal types (by factors $\sim$ 8.8, 3.6) with less significant improvement (by factors $\lesssim 2.3$) for PNG of local type and the other $\Lambda$CDM/mass cut parameters. This finding closely mirrors a previous study \cite{Coulton_2023b}, which compared constraints for the joint halo power spectrum/bispectrum with the halo power spectrum. They found that the joint halo power spectrum/bispectrum provided significantly more constraining power compared to the halo power spectrum for equilateral and orthogonal PNG, but more modest enhancements for the remaining parameters. While \cite{Coulton_2023b} used a Gaussian process to smooth derivatives to rectify the issue of convergence, and thus the results are not directly comparable to ours, they provide a qualitative framework to compare the information content of the DSC functions with the halo bispectrum. For equilateral and orthogonal PNG, there is a disproportionate amount of information encoded in the halo bispectrum compared to the halo power spectrum. Because DSC performs effectively for these two parameters in particular, we deduce that DSC is an efficient compression of higher-order information.

\subsection{Local PNG and Sample Variance Cancellation}

In the peak-background split formalism \cite{Cole_1989, Sheth_1999}, long-wavelength modes modulate the small-scale density field – the effective threshold required for halo formation changes, leading to a (constant) large-scale bias. In the presence of local PNG, the coupling between short and long-wavelength modes leads to an additional scale-dependent bias \cite{Dalal_2008}. The corresponding model for the galaxy-matter cross power spectrum (in real space) takes the following form on linear scales: 
\begin{equation} \label{eq:11} 
    P_{\rm gm}(k, \fnlloc) = (b + \Delta b (k) \fnlloc) P_{\rm mm}(k) = b\left(1+ \frac{\Delta b(k)}{b} \fnlloc\right) P_{\rm mm}(k) \,,
\end{equation} 
where 
\begin{equation} \label{eq:12} 
    \Delta b(k) \propto \frac{b_{\Phi}}{k^2} \,,
\end{equation} 
in the limit $k \rightarrow 0$ where the transfer function $T(k) \rightarrow 1$, and 
\begin{equation}\label{eq:13} 
    b_{\Phi} = \frac{\partial \ln{\bar{n}}}{\partial \ln{\sigma_8}} \,, 
\end{equation} 
measures the response of the galaxy number density to changes in the amplitude of small-scale fluctuations. For multiple tracers of the same underlying matter field with different $b_{\Phi}/b$, we would expect to find sample variance cancellation \cite{Seljak_2009}, meaning the error only depends on the shot noise. We might expect that DSC, which generates samples with different large-scale biases, would benefit from sample variance cancellation in a similar manner to halo catalogues split according to mass. However, we find that DSC provides a smaller improvement for PNG of local type (by factor $\sim 1.4$) compared to the other PNG types. If we restrict to large scales, the information gain comes almost exclusively from degeneracy breaking with other parameters; there is essentially no improvement for the unmarginalized constraints when the maximum wavenumber goes to zero, as shown in Fig.~\ref{fig:constraints_max_wavenumber}. 

We can measure the response of the DSC quantiles to $\fnlloc$ directly and check for the criteria necessary to achieve sample variance cancellation, by comparing the quantile-matter cross power spectra with different $\fnlloc$ \footnote{We use $256^3$ mesh grids and CIC interpolation for computations involving the matter fields (using 500 realizations for the fiducial, $\fnlloc$ and $\sigma_8$ variations) since matter density grids are provided in this format and regenerating them with a higher resolution was not feasible due to computational and storage limitations. This is unlikely to significantly alter results since the cell size is already small and aliasing only affects small scales.}. The two variations $\fnlloc = \pm 100$ should give opposite responses; we average the two to obtain a more robust estimate instead of relying on either variation exclusively:
\begin{equation} \label{eq:14}
    \frac{\Delta b(k) \fnlloc}{b} \equiv \frac{1}{2}\left [ {\frac{P_{\rm qm}(k, \fnlloc)-P_{\rm qm}(k, -\fnlloc)}{P_{\rm qm}(k, \fnlloc=0)}} \right ] \,.
\end{equation}
We can also calculate the (expected) response to $\fnlloc$ indirectly from Eq.~\ref{eq:13}, by measuring how the tracer density changes for simulations with different amplitude of small-scale fluctuations $\sigma_8^{\pm} = 0.834 \pm 0.015$ \cite{Castorina_2018}:
\begin{equation} \label{eq:15}
    b_{\Phi} \approx \frac{\ln{(\bar{n}(\sigma_8^+))}-\ln{(\bar{n}(\sigma_8^-))}}{\ln{(\sigma_8^+)}-\ln{(\sigma_8^-)}} = \frac{\ln{(\bar{n}(\sigma_8^+)/\bar{n}(\sigma_8^-))}}{\ln{(\sigma_8^+/\sigma_8^-)}}\,.
\end{equation}
For DSC, our `tracers' are represented by pixels in the grid corresponding to each density environment; we need to consider, for variations in $\sigma_8$, the change in the number of pixels between fixed thresholds. To choose these thresholds, we average the quantile overdensity boundaries over all fiducial simulations. Combining with a measurement of the linear bias $b$ from the large-scale ratio (averaged on scales $k \leq 0.03 \ h\mathrm{Mpc}^{-1}$) of the quantile-matter cross power spectra and the matter power spectrum in real space without local PNG, gives the (expected) ratio $b_{\Phi}/b$ for each quantile.

\begin{figure}
    \centering
    \includegraphics[width=\textwidth]{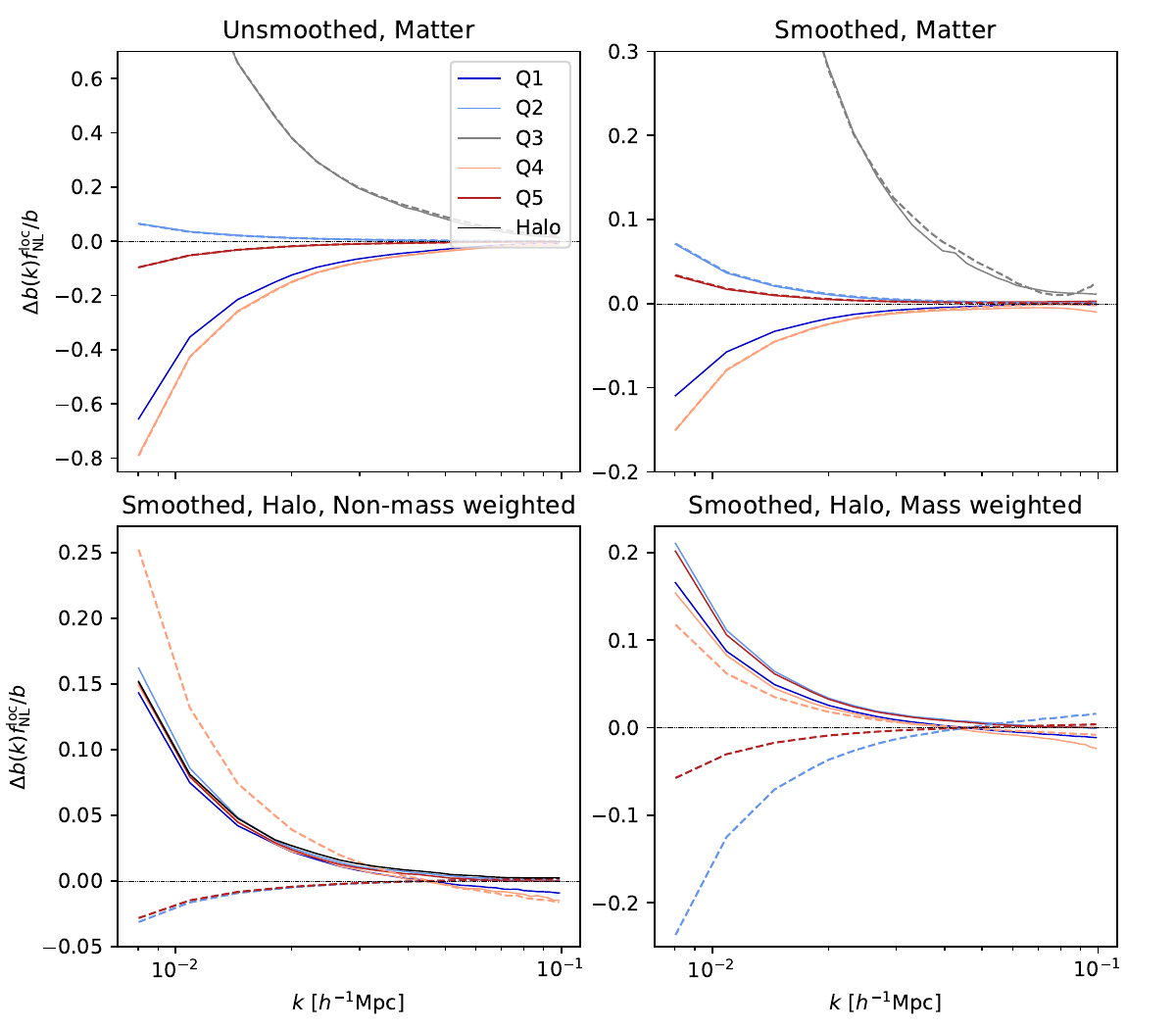}
    \caption{The observed $\fnlloc$ responses, in solid lines, for the five DSC quantiles (along with the unsmoothed non-mass-weighted halo field shown in black) according to the unsmoothed matter field, smoothed matter field (Gaussian filter radius $\Rs = 10 \ \invhMpc$), smoothed non-mass-weighted halo field (conventional DSC), and smoothed mass-weighted halo field. The dashed lines show the Q1 response re-scaled according to the relative values of $b_{\Phi}/b$ (indirectly computed from Eq.~\ref{eq:13}) for the remaining quantiles. The Q3 responses are excluded from the bottom two panels due to their comparatively large dynamical range and noise which distracted from the remaining responses and could be misinterpreted as vertical lines.}
    \label{fig:fnlloc_response}
\end{figure}

Fig.~\ref{fig:fnlloc_response} shows the observed $\fnlloc$ responses when the quantiles are defined by: the unsmoothed matter field, the smoothed matter field (with Gaussian filter radius $R_{\rm s} = 10 \ \invhMpc$), the smoothed halo number density field (not weighted by mass), and the smoothed halo field weighted by mass, and are compared with the expected responses based on the modelling of Eq.~\ref{eq:13}. The results from the matter field indicate that sample variance cancellation is happening (albeit with less strong differential responses for the smoothed field compared to the unsmoothed field), while those from the halo field indicate it is not. Furthermore, there is strong agreement between the observed and expected responses for quantiles split from the matter field whether the field is smoothed or not, versus strong disagreement when using the smoothed halo field whether weighted by mass or not. The observed responses to $\fnlloc$ in the case of the smoothed non-mass-weighted halo field (i.e., DSC as implemented in our Fisher analysis)for Q1, Q2, Q4, Q5 are nearly identical (Q3 has a different response, but the linear bias is nearly zero making the ratio noisy and unreliable), which validates the observed lack of sample variance cancellation from our Fisher forecasts. In contrast, the indirect approach predicts ratios $b_{\Phi}/b \sim$ (0.18, -0.038, -11, 0.31, -0.035) for the five quantiles, which are sufficiently different that we would expect sample variance cancellation to play a role – in theory, the relative amplitudes of the observed $\fnlloc$ responses should mirror the relative amplitudes of the ratios $b_{\Phi}/b$, but we see that they do not.

The lack of sample variance cancellation can be understood from the fact that we are deriving the DSC quantiles from a tracer field which is biased and only has a limited range of halo masses; halos below the mass cut $\mmin = 3.2 \times 10^{13} \ \solmass$ (lower biased objects) are excluded and the information regarding their $\fnlloc$ responses is not present. As the range of halo masses included increases, we expect sample variance cancellation to become increasingly strong, eventually matching what we have found when applied to the (unbiased) smoothed matter field; in the limit that halos of arbitrarily low masses are included, the smoothed mass-weighted halo field should fully characterize the smoothed matter field. In this context, weighting the halos by mass increases the range of $\fnlloc$ responses between quantiles, allowing DSC to differentiate between halos of different masses, increasing the information content.

\section{Summary}\label{sec:summary}

In this paper, we have utilized the Fisher information formalism applied to the Quijote and Quijote-PNG simulations to study the potential for using DSC to constrain PNG of local, equilateral and orthogonal types. After applying the necessary corrections for derivative convergence, we found that a joint DSC/halo power spectrum analysis provided factors of $\sim$ 1.4, 8.8, and 3.6 improvement over a halo power spectrum analysis for PNG of local, equilateral and orthogonal types. These results predict the most significant improvement for PNG of equilateral and orthogonal types, solidifying the hypothesis that DSC is an effective way of probing higher-order information missing from two-point statistics. We observed less significant improvement for PNG of local type due to the absence of significant sample variance cancellation on large scales from scale-dependent bias. We found strong agreement between the observed $\fnlloc$ response and the expected response based on the modelling of scale-dependent bias when the quantiles are defined based on either the unsmoothed or smoothed matter overdensity field. Yet, we found strong disagreement when quantiles are defined based on the smoothed halo overdensity field, even when weighted by halo mass – a field generated based on halos within a limited mass range encodes a particular $\fnlloc$ response and omits information regarding the $\fnlloc$ response to halos outside of the mass range. This is likely to limit the ability of DSC to allow sample variance cancellation for $\fnlloc$ when applied to future surveys with a single population of galaxies.

To perform our analysis, we have introduced a Fourier space DSC analysis for the first time and tested a modification of the existing algorithm by using equally spaced lattice query positions instead of randoms, which was found to significantly boost constraining power on small scales. We also briefly studied the impact of varying hyperparameters including the smoothing radius and the number of quantiles, and found that reduced smoothing radius and increased number of quantiles both boosted the information content of the DSC statistics – valuable insights for upcoming applications of DSC to achieve maximal constraining power.

\acknowledgments

All authors thank Neal Dalal for helpful discussions and William Coulton for providing Fisher matrices for comparison with our results. WP acknowledges the support of the Natural Sciences and Engineering Research Council of Canada (NSERC), [funding reference number RGPIN-2019-03908] and from the Canadian Space Agency. Research at Perimeter Institute is supported in part by the Government of Canada through the Department of Innovation, Science and Economic Development Canada and by the Province of Ontario through the Ministry of Colleges and Universities. This research was enabled in part by support provided by Compute Ontario (computeontario.ca) and the Digital Research Alliance of Canada (alliancecan.ca). We thank the anonymous referee for their insightful comments and suggestions.

\appendix

\section{Lattice/Random Query Positions} \label{ap:query_position_types}

\begin{figure}
    \centering
    \includegraphics[width=\textwidth]{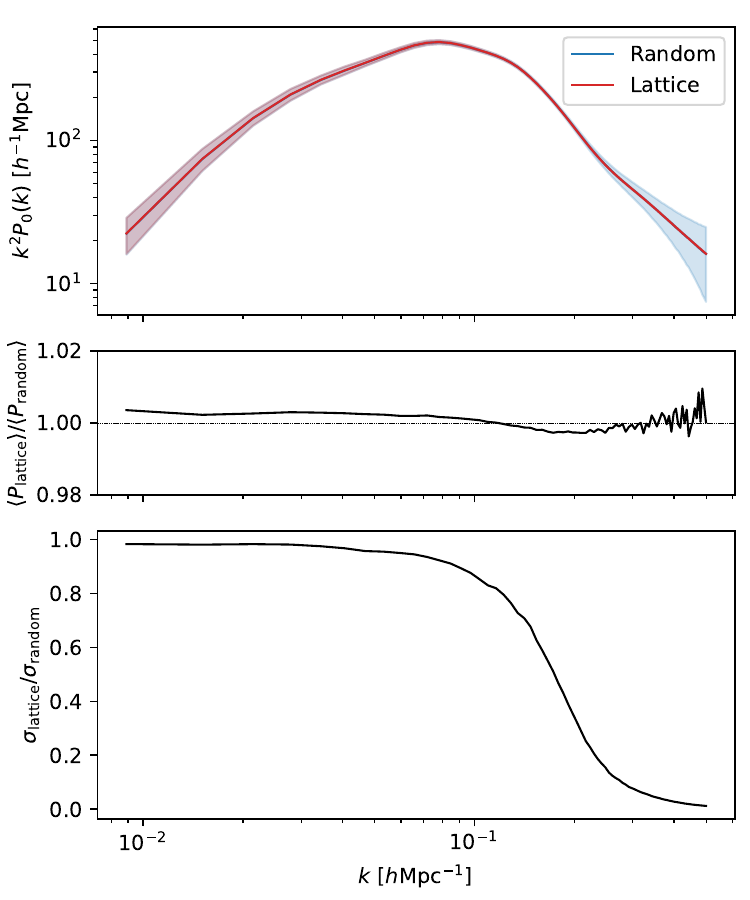}
    \caption{Quantile autocorrelation monopole for Q5 (the most overdense quantile) when using lattice or randomly positioned query positions (five times the number of halos). The top panel shows the mean power spectra and the standard noise for both methods. The middle panel shows the ratio of the mean power spectra for both versions. The lower panel shows the ratio of the standard noise between the lattice and random versions.}
    \label{fig:lattice_randoms_power}
\end{figure}

We expect the ensemble mean DSC power spectra using lattice and random query positions to agree closely. Indeed, over the range of wavenumbers probed in this study, we find only percent-level disagreement. However, we expect to see significant differences in noise between these two variations. We expect the noise to agree closely on large scales – where sample variance dominates over shot noise, but for the lattice version to reduce noise on small scales, where the additional shot noise introduced using randoms is eliminated. Fig.~\ref{fig:lattice_randoms_power} confirms both of these predictions by comparing the power spectra for the lattice and random conventions, using five times as many randoms as halos, for one of the DSC quantile autocorrelation monopoles. As shown earlier, this reduction in noise on small scales indeed translates to improved constraining power overall.

\section{Derivative Convergence Correction} \label{ap:convergence_issues}

To derive the analytic expression for the derivative convergence, we first make several simplifying assumptions. We assume that we only have one parameter being varied and one entry in our data vector, and thus the covariance matrix is replaced by the variance of the data entry. The estimator for the Fisher matrix is given by: $$\hat{F} = \frac{1}{\sigma_{\mu}^2}\left(\frac{\partial{\hat{\bar{\mu}}}}{\partial \theta}\right)^2\,.$$ Since the number of covariance realizations is much larger than the number of derivative realizations, we also assume the noise in the covariance matrix is negligible compared to the derivatives, and thus we assume the variance here is a constant. We take the expectation value of the Fisher information: $$\left < \hat{F} \right > = \frac{1}{\sigma_{\mu}^2} \left < \left ( \frac{\partial \hat{\bar{\mu}}}{\partial \theta} \right )^2 \right > = \frac{1}{\sigma_{\mu}^2} \left [\left ( \frac{\partial \bar{\mu}}{\partial \theta} \right )^2 + \text{Var} \left( \frac{\partial \hat{\bar{\mu}}}{\partial \theta} \right) \right ] = F + \frac{1}{\sigma_{\mu}^2} \text{Var} \left( \frac{\partial \hat{\bar{\mu}}}{\partial \theta} \right)\,.$$ Since $$\hat{\bar{\mu}} = \frac{1}{N_{\text{deriv}}} \sum_{\rm i}^{N_{\text{deriv}}} \mu_{\rm i}\,,$$ is the average across many derivative realizations, we can apply the central limit theorem to obtain the variance of the mean in terms of the variance of a single realization: $$\left < \hat{F} \right > = F + \frac{1}{\sigma_{\mu}^2 N_{\text{deriv}}} \text{Var} \left ( \frac{\partial \hat{\mu}_{\rm i}}{\partial \theta} \right )  = F \left( 1 + \frac{1}{\sigma_{\mu}^2 N_{\text{deriv}} F} \text{Var} \left ( \frac{\partial \hat{\mu}_{\rm i}}{\partial \theta} \right )\right)\,.$$ We can absorb the terms $\sigma_{\mu}^2$, $F$ and $\text{Var} \left ( \frac{\partial \hat{\mu}_{\rm i}}{\partial \theta} \right )$ into a single constant since we are fitting a model to the observed convergence. We therefore have: $$\left < \hat{F} \right >^{-1} = F^{-1} \left( 1 + \frac{500C}{N_{\text{deriv}}} \right )^{-1}\,,$$ where $$C \equiv \frac{\text{Var} \left ( \frac{\partial \hat{\mu}_{\rm i}}{\partial \theta} \right )}{500 \sigma_{\mu}^2 F}\,.$$ Next, we substitute the parameter errors based on the Cramér-Rao bound: $$\sigma_{\theta}^2(N_{\text{deriv}}) = \sigma_{\theta}^2 \left( 1 + \frac{500C}{N_{\text{deriv}}} \right )^{-1}\,.$$ Now, we normalize to the maximum number of realizations and take the square root: $$\frac{\sigma_{\theta}(N_{\text{deriv}})}{\sigma_{\theta}(N_{\text{deriv}}=500)} = \left(\frac{1+C}{ 1 + \frac{500C}{N_{\text{deriv}}}}\right)^{1/2}\,.$$ Although the simplifying assumptions we have made here do not hold under a more general Fisher analysis with multiple parameters and entries in the data vector, we expect the same dependence on the number of realizations to hold and we have verified that the functional form provides strong qualitative agreement with the shape of the derivative convergences.

\printbibliography
\end{document}